\documentclass[journal]{IAENGtran}

\usepackage{cite}

\ifCLASSINFOpdf
   \usepackage[pdftex]{graphicx}
   \DeclareGraphicsExtensions{.pdf,.jpeg,.png}
\else
   \usepackage[dvips]{graphicx}
   \DeclareGraphicsExtensions{.eps}
\fi

\usepackage[cmex10]{amsmath}
\usepackage{array}

\usepackage{mdwmath}
\usepackage{mdwtab}

\usepackage{url}

\newtheorem{theorem}{Theorem}
\newtheorem{lemma}{Lemma}
\def\ds{\displaystyle}
\def\e{\epsilon}
\def\RR{\hbox{I\kern-.2em\hbox{R}}}
\def\NN{\hbox{I\kern-.2em\hbox{N}}}
\newcommand{\norme}[2]{\ensuremath{\| #1 \|_{#2}}}
\providecommand{\myceil}[1]{\ensuremath{\left \lceil #1 \right \rceil }}

\begin{document}

\title{A Posteriori Error Estimator for a Front-Fixing Finite Difference Scheme for American Options}

\author{Riccardo~Fazio
\thanks{Manuscript received March 15, 2015. 
This work was supported in part by the University of Messina trhought the S.T.I.G.A.F.F. project and by the GNCS of INDAM.}
\thanks{R. Fazio is an associate professor of the Department of Mathematics and Computer Science, University of Messina, Viale F. Stagno D'Alcontres, 31 -- 98166 Messina, Italy. (work phone: +390906765064; fax: +39090393502; e-mail: {rfazio@unime.it}; home-page: {http://mat521.unime.it/fazio}). He is also a member of Finance \& Risk Lab S.r.l., see the {www.financerisklab.it} web site.}}

\maketitle

\pagestyle{empty}
\thispagestyle{empty}

\begin{abstract}
For the numerical solution of the American option valuation problem, we provide a script written in MATLAB implementing an explicit finite difference scheme. 
Our main contribute is the definition of a posteriori error estimator for the American options pricing which is based on Richardson's extrapolation theory.
This error estimator allows us to find a suitable grid where the computed solution, both the option price field variable and the free boundary position, verify a prefixed error tolerance.
\end{abstract}

\begin{IAENGkeywords}
American options, free boundary problem, finite difference scheme,  Richardson's extrapolation, a posteriori error estimator. 
\end{IAENGkeywords}

\IAENGpeerreviewmaketitle

\section{Introduction}
In the market for financial derivatives, the most important problem is the so-called {\it option valuation problem} or in a few words: the problem of computing a fair price for a given option. 
An American call (put) option is a contract written on an underlying asset and gives the holder the right to buy (sell) the asset for a pre-specified price  or strike price on or before a pre-specified date also indicated as maturity. 
Unlike European options where the holder can exercise the option only on the maturity date, the possibility of early exercise makes the pricing of American options a problem in stochastic optimization. 
While a closed-form solution for the price of European options is derived in the celebrated work by Black and Scholes \cite{Black:POC:1973} and by Merton \cite{Merton:1973:TRO}, there exists no analogous result for American options. 
The reason can be explained as follows: while the governing differential equation is still the one obtained by Black and Scholes \cite{Black:POC:1973}, McKean \cite{McKean:1965:AFB} and Merton \cite{Merton:1973:ICA} show that the price of an American option satisfies boundary conditions governed by a boundary that is not known a priori and needs to be computed as part of the solution itself. 
Such problems are called free-boundary problems. 
In particular, the American call option problem is a free boundary problem defined on a finite interval.
On the other hand, the American put option problem is a free boundary problem defined on a semi-infinite interval so that it is a non-linear problem complicated by a boundary condition at infinity.
Therefore, both such derivatives of financial markets must be priced by numerical or analytical approximations.

Within analytical approximations, MacMillan \cite{MacMillan:1986:AAA} and Barone-Adesi and Whaley \cite{Barone-Adesi:1987:EAA} defined a quadratic approximation  for American put option.
These methods are not convergent and have trouble pricing long-maturity options accurately. 
To correct this problem, Ju and Zhong (1999) develop an approximation based on the method proposed in Barone-Adesi and Whaley.
While this improved method prices long-maturity options more accurately, it is still not convergent.
Johnson \cite{Johnson:1983:AAA} used an interpolation scheme to price the American put option, Geske and Johnson \cite{Geske:1984:APO} derived a valuation formula expressed in terms of a series of compound-option functions for the same reason, 
Bunch and Johnson \cite{Bunch:1992:SNE} propose a modified two-point Geske-Johnson approach.
Carr and Faguet \cite{Carr:1996:VFL} view put options as the limit of a sequence of perpetual option values which are subject to default risk, and use this view to deriving approximations for the price of an option.
More recently, Zhu \cite{Zhu:2006:EES} derives a semi-closed form solution for the price of the option as a Taylor series expansion consisting of infinite terms, but requiring thirty terms for an accurate option price.

Within numerical approximations, the most popular methods for pricing American option can be classified to lattice method, Monte Carlo simulation, and Finite difference method.
The lattice method was first introduced by Cox et al. \cite{Cox:1979:OPS}, and its convergence was proved by Amin and Khanna \cite{Amin:1994:CAO}.
Fu \cite{Fu:1994:OUS,Fu:1994:TRT} applied Monte Carlo method along with gradient-based optimization techniques to price American style options.
The application of a finite difference method to price American options was initiated by Brennan and Schwartz \cite{Brennan:1977:VAP,Brennan:1978:FDM} and Schwartz \cite{Schwartz:1977;VWI}.
Jaillet et al. \cite{Jaillet:1990:VIP} proved the finite difference method convergence.
A front-fixing finite difference method was proposed by Wu and Kwok (1997), Nielsen et al. \cite{Nielsen:2002:PFF}, and Company et al. \cite{Company:2014:SAP} to compute option prices.
The front-fixing method utilizes a change of variables to transform the free boundary problem into a nonlinear problem on a fixed domain. 
Nielsen et al. \cite{Nielsen:2002:PFF} also propose a penalty method to price American put options, where the unknown boundary is removed by adding a penalty term, again leading to a nonlinear problem posed on a fixed domain.

In this paper, we list a script written in MATLAB implementing a finite difference scheme for the numerical solution of the American option models of financial markets.  
We implemented both the method defined by Company et al. \cite{Company:2014:SAP} and the method developed by Nielsen et al. \cite{Nielsen:2002:PFF} and found that the first method implementation is more efficient than the second one.
Our main contribute is the definition of a posteriori error estimator for the American options pricing which is based on Richardson's extrapolation theory.
This error estimator allows us to find a suitable grid where the computed solution, both the option price field variable and the free boundary position, verify a prefixed error tolerance. 

\section{American put option}
Let us suppose that at time $t$ the price of a given underlying asset is $S$.
We consider here the following mathematical model for the value $ P = P(S, \tau)$ of an American put option to sell the asset:
\begin{align}
& {\ds \frac{\partial P}{\partial \tau}} = \frac{1}{2} \sigma^2 S^2  
{\ds \frac{\partial^2 P}{\partial S^2}} + r S {\ds \frac{\partial P}{\partial S}} -r P \ , \nonumber \\[1ex]
&  P(S, 0) = \max(E-S, 0) \ , \qquad S^*(0) = E \ ,  \nonumber \\[1ex]
& \lim_{S\rightarrow \infty} P(S, \tau) = 0 \ , \label{APO:model} \\
& P(S^*(\tau), \tau) = E - S^*(\tau) \ , \qquad {\ds \frac{\partial P}{\partial S}}(S^*(\tau), \tau) = 
- 1 \ , \nonumber \\
& P(S, \tau) = E-S \ , \quad 0 \le S < S^*(\tau) \nonumber
\end{align}
where $\tau = T-t$, $t$ denotes the time to maturity $T$, $S^*(\tau)$ is a free boundary, that is the unknown early exercise boundary, $\sigma$, $r$ and $E$ are given constant parameters representing the volatility of the underlying asset, the interest rate and the exercise price of the option, respectively, and the governing equation is defined on $0 \le \tau \le T$, $S^*(\tau) < S < \infty$.

To fix the free boundary we apply the dimensionless new variables 
\begin{align}\label{eq:front-fixing}
x &= \ln \frac{S}{S_f(\tau)} \ , \nonumber \\
S_f &= \frac{S^*(\tau)}{E}\ , \\
p(x, \tau) &= \frac{P(x S_f(\tau), \tau)}{E} \ , \nonumber
\end{align}
see Wu and Kwok \cite{Wu:1997:FFM}.
From the variables transformation defined by (\ref{eq:front-fixing}) follows that $S_f(\tau)$ is mapped on the fixed line $x=0$, $0 \le p(x, \tau) \le 1$ and $0 \le S_f(\tau) \le 1$.  
By using (\ref{eq:front-fixing}), the put option problem (\ref{APO:model}) can be rewritten as follows
\begin{align}
\label{APO:model:fix}
& {\ds \frac{\partial p}{\partial \tau}} = \frac{1}{2} \sigma^2 
{\ds \frac{\partial^2 p}{\partial x^2}} + \left(r - \frac{\sigma^2}{2}\right) {\ds \frac{\partial p}{\partial x}} -r p + \frac{1}{S_f(\tau)} \frac{dS_f}{d\tau}(\tau) {\ds \frac{\partial p}{\partial x}}   \ , \\[1ex]
\label{APO:model:fix1}
&  p(x, 0) = 0 \quad \mbox{for} \quad 0 \le x \ , \qquad S_f(0) = 1 \ ,  \\[1ex]
\label{APO:model:fix2}
& \lim_{x \rightarrow \infty } p(x, \tau) = 0 \ , \\
\label{APO:model:fix3}
& p(0, \tau) = 1 - S_f(\tau) \ , \qquad {\ds \frac{\partial p}{\partial x}}(0, \tau) = 
- S_f(\tau)  \ , 
\end{align}
that has to be solved on the domain defined by $0 \le \tau \le T$ and $0 < x < \infty$.

\section{An explicit finite difference scheme}
To solve the problem (\ref{APO:model:fix}-\ref{APO:model:fix3}) numerically, we introduce a truncated boundary $x_\infty$, which is a suitable large value where it would be convenient to impose the asymptotic boundary condition.
In other words, we replace the asymptotic boundary condition (\ref{APO:model:fix2}) with the side condition
\begin{equation}\label{eq:FBC} 
p(x_\infty, \tau) = 0 \ .
\end{equation}
For the choice of  $x_\infty$ and the accuracy of the related numerical solution, we can refer to the study by Kangro and Nicolaides \cite{Kangro:2000:FFB}.
On the other hand, the boundary condition at infinity can be enforced exactly by using a non-standard finite difference scheme, and this has been shown for the numerical solution of the so-called perpetual American put option in \cite{Fazio:201?:PAP}.

Next, by setting an integer $J$ and a positive value $\mu$, we define the step-sizes
\begin{equation}\label{eq:step-sizes} 
\Delta x=\frac{x_\infty}{J} \ , \quad \Delta t = {\mu}{\Delta x}^{2} \ ,
\end{equation}
the integer $N$ 
\begin{equation}\label{eq:N} 
N = \myceil{\frac{T}{\Delta t}} \ ,
\end{equation}
where $ \lceil\cdot\rceil : \RR^+ \rightarrow \NN $ is the \textit{ceiling} function which maps a real number to the least integer that is greater than or equal to that number.
Therefore, $\mu$ is the grid-ratio 
\begin{equation}\label{eq:c}
\mu = \frac{\Delta t}{\Delta x^2} \ .
\end{equation}
So that, within the finite domain, we can introduce a mesh of grid-points 
\begin{equation}\label{eq:grid} 
x_j = j \Delta x \ , \quad t_n = n \Delta t \ ,
\end{equation}
for $j =0, 1, \dots , J$ and $n = 0, 1, \dots, N$. 

We would like to define a numerical scheme that allows us to compute the grid values
\begin{equation}\label{eq:p}
p_j^n \approx p(x_j,t_n) \ , 
\end{equation}
for $j =0, 1, \dots , J$ and $n = 0, 1, \dots, N-1, N$ and the free boundary values
\begin{equation}\label{eq:R} 
S_f^n \approx S_f(t_n) \ ,
\end{equation}
for $n = 0, 1, \dots, N-1, N$.
To this end let us consider the explicit finite difference scheme
\begin{align}\label{eq:MN1}
\frac{p_j^{n+1}-p_j^n}{\Delta t} &= \frac{1}{2} \sigma^2 \frac{p_{j-1}^{n}-2 p_j^{n}+p_{j+1}^{n}}{(\Delta x)^2} + \nonumber \\
&+ \left(r - \frac{\sigma^2}{2}\right)\frac{p_{j+1}^{n}-p_{j-1}^{n}}{2 \Delta x} + \\
&+ \frac{S_f^{n+1}-S_f^{n}}{\Delta t S_f^{n}}\frac{p_{j+1}^{n}-p_{j-1}^{n}}{2 \Delta x}-r p_j^{n} \nonumber
\end{align}
for $j =1, 2, \dots , J-1$ and $n = 0, 1, \dots, N-1$. 
For our specific problem $p_j^{n}$ and $S_f^{n}$ are given and our goal is to compute $p_j^{n+1}$ and $S_f^{n+1}$.
If we apply some simple algebra, then we can rewrite the explicit finite difference scheme as
\begin{equation}\label{eq:FDE}
p_j^{n+1} = a p_{j-1}^{n}+b p_{j}^{n}+c p_{j+1}^{n} + \frac{S_f^{n+1}-S_f^{n}}{\Delta t S_f^{n}} \frac{p_{j+1}^{n}-p_{j-1}^{n}}{2 \Delta x} \ ,
\end{equation}
for $j =2, 3, \dots , J-1$ and $n = 0, 1, \dots, N-1$, where
\begin{align}\label{eq:FD:abc}
a &= \frac{\mu}{2} \left[ \sigma^2 - \left(r - \frac{\sigma^2}{2}\right) \Delta x \right] \nonumber \\
b &= 1 - \mu \sigma^2 - r \Delta t \\
c &= \frac{\mu}{2} \left[ \sigma^2 - \left(r + \frac{\sigma^2}{2}\right) \Delta x \right] \nonumber \ .
\end{align}
Now, we have to take into account the side conditions.
From the two initial conditions (\ref{APO:model:fix1}), we obtain
\begin{equation}\label{eq:FD:fix1}
p_j^0 = 0 \ , \quad S_f^0 = 1 \ ,
\end{equation}
for $j= 0,1, \dots , J$.
From the boundary conditions (\ref{eq:FBC}), we get
\begin{equation}\label{eq:FD:fix2}
p_J^{n} = 0  \ ,
\end{equation}
for $n= 0,1, \dots , N$.
From the two boundary conditions (\ref{APO:model:fix3}), using a central finite difference formula, we derive
\begin{equation}\label{eq:FD:fix3}
\frac{p_1^n-p_{-1}^n}{2 \Delta x} = - S_f^n \ , \quad p_0^n = 1-S_f^n \ ,
\end{equation}
where $x_{-1} = -\Delta x$ is a fictitious point out of the computational domain.
Moreover, by considering the governing differential equation (\ref{APO:model:fix}) at $x_0 = 0$, $\tau > 0$ and taking into account the side conditions (\ref{APO:model:fix3}) one gets a new boundary condition:
\begin{equation}\label{APO:model:fix5}
\frac{\sigma^2}{2}\frac{\partial^2 p}{\partial x^2}(0^+, \tau) + \frac{\sigma^2}{2} S_f(\tau) -r = 0 \ ,
\end{equation}
see Wu and Kwok \cite{Wu:1997:FFM}, Zhang and Zhu \cite{Zhang:2009:HFD} or Kwok \cite[p. 341]{Kwok:2008:MMF} , and its central finite difference discretization
\begin{equation}\label{eq:FD:fix5}
\frac{\sigma^2}{2}\frac{p_{-1}^{n}-2 p_{0}^{n}+p_{1}^{n}}{\Delta x^2} + \frac{\sigma^2}{2} S_f^n -r = 0 \ .
\end{equation}
Now, we can eliminate the value of $p_{-1}^{n}$ from equations (\ref{eq:FD:fix3}) and (\ref{eq:FD:fix5}) to get
\begin{equation}\label{eq:FD:fix6}
p_{1}^{n} =  1 + \frac{r \Delta x^2}{\sigma^2}- \left(1 + \Delta x + \frac{\Delta x^2}{2}\right) S_f^n \ .
\end{equation}
If we evaluate the numerical scheme (\ref{eq:FDE}) for $j=1$ and take into account (\ref{eq:FD:fix6}) for $n = n+1$, then the free boundary $S_j^{n+1}$ can be defined by
\begin{equation}\label{eq:FDSf}
S_f^{n+1} = d^{n} S_f^{n} \ .
\end{equation}
for $n = 0, 1,\dots , N-1$, where 
\begin{equation}\label{eq:FD:dn}
d^{n} = \frac{1 + {\ds \frac{r \Delta x^2}{\sigma^2}}- \left(a p_0^n + b p_1^n + c p_2^n - {\ds \frac{p_2^n - p_0^n}{2 \Delta x}}\right)}{{\ds \frac{p_2^n - p_0^n}{2 \Delta x}} + \left(1 + \Delta x + {\ds \frac{\Delta x^2}{2}}\right) S_f^n} \ .
\end{equation}
We are now ready to define the algorithm:
\begin{enumerate}
\item Input $\sigma$, $r$, $E$, $T$, $J$, $\mu$ and $x_\infty$;
\item Define the grid ($x_j$, $\Delta x$, $t_n$, $\Delta t$) according to equations (\ref{eq:step-sizes}) and (\ref{eq:grid});
\item for $j=0, 1, 2, \dots, J$ do $p_j^{0} = 0$ end, set $S_f^{0} = 1$;
\item compute $a$, $b$ and $c$ according to (\ref{eq:FD:abc});
\item for $n=0, 1, \dots N-1$, compute $d^n$ according to (\ref{eq:FD:dn}), compute $S_f^{n+1}$ according to (\ref{eq:FDSf}) and apply the free boundary conditions
\begin{equation*}
p_0^{n+1} = 1-S_f^{n+1} \ ,
\end{equation*}
\begin{equation*}
p_{1}^{n+1} =  1 + \frac{r \Delta x^2}{\sigma^2}- \left(1 + \Delta x + \frac{\Delta x^2}{2}\right) S_f^{n+1} \ ,
\end{equation*}
compute 
\begin{equation*}
a^n = a - \frac{S_f^{n+1} - S_f^{n}}{2 \Delta x S_f^{n}} \ , \quad c^n = c - \frac{S_f^{n+1} - S_f^{n}}{2 \Delta x S_f^{n}}
\end{equation*}
and for $j = 2, 4, \dots, J-1$ compute the values of $p_j^{n+1}$ according to
\begin{equation*}
p_j^{n+1} = a^n p_{j-1}^{n}+b p_{j}^{n}+c^n p_{j+1}^{n} \ .
\end{equation*}
end, set $p_{J}^n = 0$, end.
\end{enumerate}
The implementation of this algorithm in MATLAB has been used to get the numerical results reported below.
The script file is listed in the Appendix.

\section{Positivity, monotonicity and stability}
In this section, we recall the theoretical results that make the explicit difference scheme suitable for doing numerical studies of the American put option.
For the sake of simplicity, let us define the numerical solution vector at time level $t_n$ as $p^n= (p_0^n$, $p_1^n$,$\dots$, $p_J^n)^T$.
\begin{lemma}\label{lemma1} (Due to Company et al. \cite{Company:2014:SAP}.) 
If $\Delta t$ and $\Delta x$ verify the two inequalities
\begin{align}
\label{dis:Delta1}
\Delta x &\le \frac{\sigma^2}{|r-\sigma^2/2|} \Delta t  \ , \quad r \ne \sigma^2/2 \\[-1ex]
& \nonumber \\[-1ex]
\label{dis:Delta2}
\Delta t &\le \frac{\Delta x^2}{\sigma^2+r \Delta x^2} \ . 
\end{align}
then the coefficients $a$, $b$ and $c$ are non-negative. 
If $r = \sigma^2/2$, then the non-negativity of these coefficients is verified under the condition (\ref{dis:Delta2}).
\end{lemma}
These two inequalities follow from positivity preserving considerations related to the explicit finite difference scheme.
The theoretical framework is similar to the one used by Friedrichs \cite{Friedrichs:SHD:1954} to study positivity preserving finite difference schemes for the advection equation, see also Fazio and Jannelli \cite{Fazio:2009:SOP}.
Moreover, the coefficients of $p_i^{n+1}$, for $i=j-1$,$j$,$j+1$, in our difference scheme (\ref{eq:FDE})
might be regarded, in the limit for $\Delta t \rightarrow 0$, as probability values.
In fact, they sum up to $1-r \Delta t$ and by imposing their positivity we get the two inequalities (\ref{dis:Delta1})-(\ref{dis:Delta2}). 
\begin{theorem}\label{teorema1} (Due to Company et al. \cite{Company:2014:SAP}.) 
Let $\{p_j^n, S_f^n\}$ be the computed numerical solution and $d^n$ be defined by equations (\ref{eq:FD:dn}), then under hypothesis of the Lemma \ref{lemma1}, for sufficiently small values of $\Delta x$, we have:
\begin{itemize}
\item $\{S_f^n\}$ is positive and non-increasing monotone for $n = 0, 1, \dots , N$;
\item the vectors $p^n$ have positive components for $n = 0, 1, \dots , N$;  
\item the vectors $p^n$ are non-increasing monotone with respect to the $j$ for each fixed $n = 0, 1, \dots , N$.
\end{itemize}
\end{theorem}

As far as the stability of the explicit finite difference scheme is concerned we can introduce the definition: the numerical scheme is said to be $\norme{\cdot}{\infty}$-stable if, for every mesh in the computational domain $[0, x_\infty] \times [0, T]$, there exists a positive constant $C$ such that
\begin{equation}
\norme{p^n}{\infty} \le C \qquad \mbox{for} \ \  n = 0, 1, \dots , N \ ,
\end{equation}
where $C$ is independent on $\Delta t$, $\Delta x$ and $n$, see Company et al. \cite{Company:2012:CSN}.  

\begin{theorem}\label{teorema2} (Due to Company et al. \cite{Company:2014:SAP}.) 
Under the hypothesis of Theorem \ref{teorema1} the explicit finite difference scheme for the fixed domain problem (\ref{APO:model:fix}-\ref{APO:model:fix3}) is
$\norme{\cdot}{\infty}$-stable.
\end{theorem}

\section{A posteriori error estimator}
For a scalar $U$ of interest, either a value of the solution $p_j^n$ or a free boundary value $S_f^n$, the numerical error $e$ can be defined by
\begin{equation}\label{eq:GE}
e = u - U \ ,
\end{equation}
where $u$ is the exact, usually unknown, value.
When the numerical error is caused prevalently by the discretization error and in the case of smooth enough solutions the global error can be decomposed into a sum of powers of the inverse of $N$ 
\begin{equation}\label{eq:asymE}
u = U_N + C_0 \left(\frac{1}{N}\right)^{p_0}+ C_1 \left(\frac{1}{N}\right)^{p_1}+ C_2 \left(\frac{1}{N}\right)^{p_2}+ \cdots \ ,
\end{equation}
where $C_0$, $C_1$, $C_2$, $\dots$ are coefficient that depend on $u$ and its derivatives, but are independent on $N$, and $p_0$, $p_1$, $p_2$, $\dots$ are the true orders of the discretization error, see Schneider and Marchi \cite{Schneider:GRR:2005} and the references quoted therein.
Each $p_k$ is usually a positive integer with $p_0 < p_1 < p_2 < \cdots$ and all together they constitute an arithmetic progression of ratio $p_1-p_0$.
The value of $p_0$ is called the asymptotic order or the order of accuracy of the method or of the numerical value $U_N$. 
By replacing into equation (\ref{eq:asymE}) $N = N_g$ and $N = N_{g+1}$ and subtracting, to the second obtained equation the first times $(1/q)^{p_0}$, $q = N_{g+1}/N_{g}$, we get the first extrapolation formula 
\begin{equation}\label{eq:Rextra1}
u \approx  U_{g+1} + \frac{U_{g+1}-U_{g}}{q^{p_0}-1} \ ,
\end{equation}
that has a leading order of accuracy equal to $p_1$.
This type of extrapolation is due to Richardson \cite{Richardson:1910:DAL,Richardson:1927:DAL}.
Taking into account equation (\ref{eq:Rextra1}) we can conclude that the error estimator by a first Richardson's extrapolation is given by
\begin{equation}\label{eq:est1}
e_{r} = \frac{U_{g+1}-U_{g}}{q^{p_0}-1} \ ,
\end{equation}
where $p_0$ is the order of the numerical method used to compute the numerical solutions.
Hence, (\ref{eq:est1}) gives the real value of the numerical solution error without knowledge of the exact solution.
In comparison with (\ref{eq:est1}) a safer error estimator can be defined by
\begin{equation}\label{eq:est2}
e_{s} = U_{g+1}-U_{g} \ .
\end{equation}
Of course, $p_0$ can be estimated with the formula
\begin{equation}\label{eq:p0}
p_0 \approx {\ds \frac{\log(|U_g-u|)-\log(|U_{g+1}-u|)}{\log(q)}} \ ,
\end{equation}
where $u$ is again the exact solution (or, if the exact solution is unknown, a reference solution computed with a suitable large value of $N$), and both $u$ and $U_{g+1}$ are evaluated at the same grid-points of $U_g$. 

Within the above framework, in order to improve the numerical accuracy by using only a small number of grid-nodes, we can generalize (\ref{eq:Rextra1}) introducing the following repeated extrapolation formula
\begin{equation}\label{eq:Rextra}
U_{g+1,k+1} = U_{g+1,k} + \frac{U_{g+1,k}-U_{g,k}}{q^{p_k}-1} \ ,
\end{equation}
where $g \in \{0, 1, 2 , \dots , G-1\}$, $k \in \{0, 1, 2, \dots , G-1\}$, $q = N_{g+1}/N_{g}$ is the grid refinement ratio, and $p_k$ is the true order of the discretization error.
The formula (\ref{eq:Rextra}) is asymptotically exact in the limit as $N_0$ goes to infinity if we use uniform grids.
We notice that to obtain each value of $U_{g+1,k+1}$ requires two computed solutions $U$ in two adjacent grids, namely $g+1$ and $g$ at the extrapolation level $k$.
For any $g$, the level $k=0$ represents the numerical solution of $U$ without any extrapolation.
We recall that the theoretical orders of accuracy of the numerical values $U_{g,k}$, with $N =N_g$ and $k$ extrapolations, verify the relation
\begin{equation}\label{eq:pk}
p_k = p_0 + k (p_1-p_0) \ ,
\end{equation}
where this equation is valid for $k \in \{0, 1, 2, \dots , G-1\}$.

\section{Numerical results}
We introduce a numerical test for the finite difference schemes defined below.
To this end, we consider the American put option problem (\ref{APO:model:fix}-\ref{APO:model:fix3}) with the following parameters:
\begin{equation}\label{eq:parameters}
r = 0.1 \ , \quad \sigma = 0.2 \ , \quad E = T = 1  \ .
\end{equation}

First of all, we are interested to define a suitable value of the truncated boundary $x_\infty$.
Then, we investigate numerically how the choice of the value of $x_\infty$ influences the numerical solution. 
To take a simple approach, we can monitor the final free boundary computed values $S_f^N$. 
Some sample results are reported in Table~\ref{tab:Rxinf} and indicate that we can set $x_\infty =1$. 

\begin{table}[!hb]
\renewcommand\arraystretch{1.3}
\caption{Free boundary location at $t=T$.}
\centering
\begin{tabular}{cccc}
\hline
{$x_\infty$} & {$N = 10$} & {$N = 20$} & {$N = 40$} \\
\hline  
$1$  & $0.8676354534435$  & $0.8655750222427$ & $0.86438663362975$  \\ 
$2$  & $0.8676354534435$  & $0.8655750222427$ & $0.86438663362975$  \\ 
$4$  & $0.8676354534435$  & $0.8655750222427$ & $0.86438663362975$   \\ 
\hline
\end{tabular}
\label{tab:Rxinf}
\end{table}

\begin{figure}[!t]
\begin{picture}(300,200)(0,0)
\put(1,1){\resizebox{9cm}{!}{\includegraphics{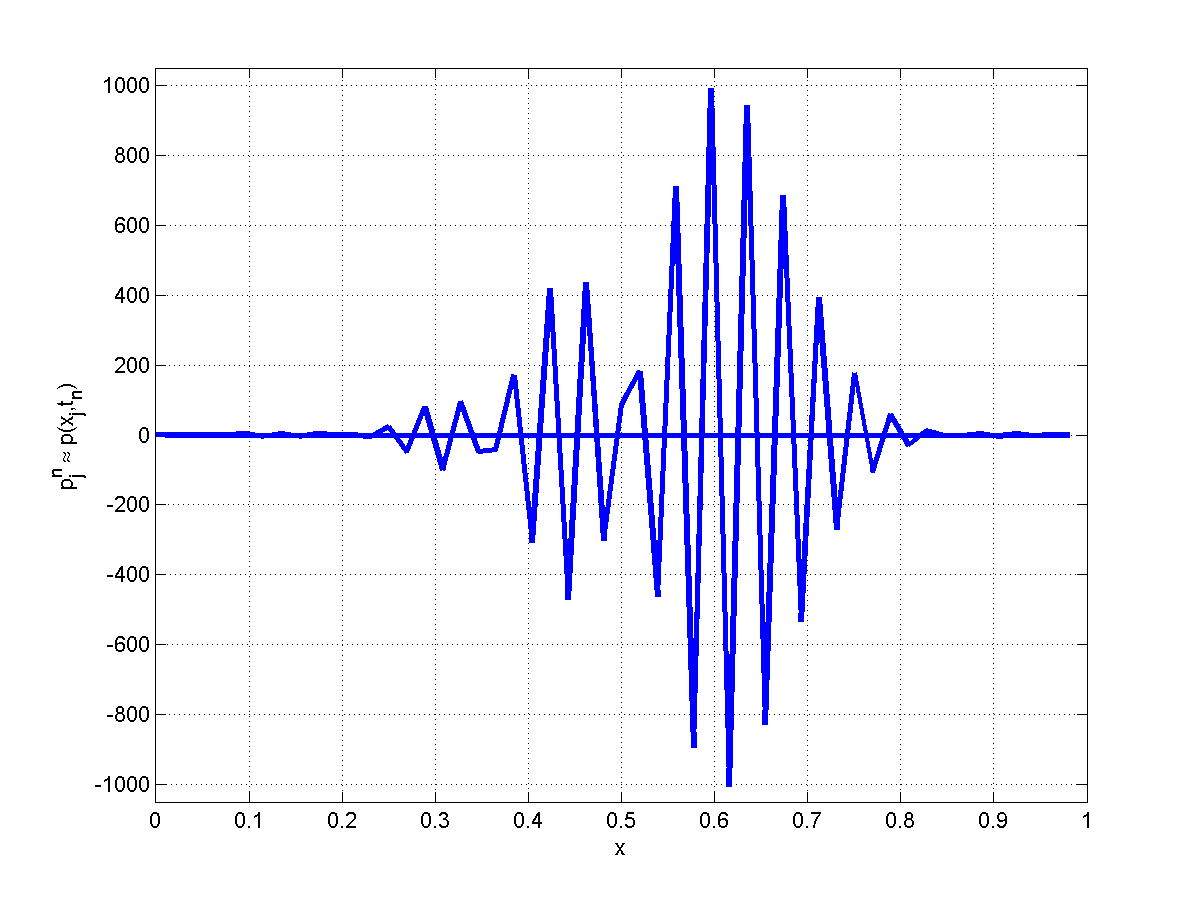}}}
\end{picture}
\vspace{-.3cm}
\caption{\it Instability for $N = 100$, $\mu =27$ and $x_\infty = 1$.}
\label{fig:fdAPOin}
\end{figure}

Fig.~\ref{fig:fdAPOin} shows an unstable computation.
For the sake of clarity, in Fig.~\ref{fig:fdAPOin} we display only the initial condition and the numerical solution at $t=T$, that is the last time iterate.
As usual, the instability manifest itself with large oscillations between positive and negative values.
Therefore, as far as our explicit finite difference scheme is concerned, we have to find a compromise between accuracy and stability.
Of course, from now on, the chosen grid-spacings are defined in order to verify the stability inequalities (\ref{dis:Delta1})-(\ref{dis:Delta2}).

From the results listed in Table~\ref{tab:Rxinf} we realize that, fixed a value of the truncated boundary $x_\infty$, the computed values of $S_f^N$ for different values of the grid-steps are in agreement only for the first two decimal places.
Then, we decided to improve the numerical accuracy by performing a mesh refinement.
Moreover, we applied repeated Richardson's extrapolation to improve the numerical accuracy.
Let us recall that the explicit difference scheme is first order in time and second order in space both for the field variable and the free boundary value, i.e. the truncation error is of the order $O(\Delta t^2) + O(\Delta t \Delta x^2)$.
We will use this result below when we perform a mesh refinement keeping constant the grid-ratio, i.e. $\mu$, so that we end up with second-order truncation error $T_j^n = O(\Delta t^2)$ in time and, therefore, the global error is first-order, that is the $p_0$ value defined above is equal to one.
We remark that in our case the sequence of $q^{p_k}$, for $k=0,1, \dots$, is given by $4$, $16$, $64$, $256$, $1024$, $\dots$, that is $q =4$ and $p_k = k+1$, for $k=0,1, \dots$. 
In Table~\ref{tab:Rextra} we report sample numerical results for the benchmark value $S_f^N$.
\begin{table}[!hbt]
\renewcommand\arraystretch{1.2}
\caption{Richardson's repeated extrapolations for the free boundary value at $t=T$.
.}\label{tab:Rextra} 
\centering
\begin{tabular}{r|ccccc}
\hline
{$N$} & {$U_{g,0}$} & {$U_{g,1}$} & {$U_{g,2}$} & {$U_{g,3}$} & {$U_{g,4}$} \\
\hline  
$5$            & $0.871621$ & $$ & $$ & $$ & $$ \\
$20$          & $0.865575$ & $0.863560$ & $$ & $$ & $$ \\
$80$          & $0.863700$ & $0.863075$ & $0.863043$ & $$ & $$ \\
$320$        & $0.863071$ & $0.862861$ & $0.862847$ & $0.862844$ & $$ \\
$1280$      & $0.862859$ & $0.862788$ & $0.862783$ & $0.862782$ & $0.862782$ \\
$5120$      & $0.862788$ & $0.862764$ & $0.862763$ & $0.862762$ & $0.862762$ \\
\hline
\end{tabular}
\end{table}
Within the same Table, we report the results obtained by repeated Richardson's extrapolations. 
We remark that the next extrapolated value from Table~\ref{tab:Rextra} is $U_{5,5}=0.862762$ so that our benchmark value is $S_f^{N} \approx 0.862762$.
\ This value can be compared with the values $S_f^{N} \approx 0.86222$ computed by Nielsen et al. \cite{Nielsen:2002:PFF} and $S_f^{N} \approx 0.8628$ found by Company et al. \cite{Company:2014:SAP}.

Next, we indicate how to use the error estimator defined by equation (\ref{eq:est1}), or alternatively by equation (\ref{eq:est2}).
Let us assume that our goal is to solve the American put option problem with a given tolerance $\e$, where $0 < \e \ll 1$. 
To this end we should solve the given problem twice, for two grids defined with given values of $J_g = J$ and $J_{g+1} =2 J$ of space intervals but for the same value of the grid-ratio $\mu$.
The corresponding time intervals $N_g$ and $N_{g+1}$ verify the relation $q = N_{g+1}/N_{g}$.
Hence we can apply (component-wise) to $p^n$ and to $S_f^n$ the error estimator formula (\ref{eq:est1}), or (\ref{eq:est2}). 
Then, we can verify whether, for $n = 1, 2, \dots, N$,
\begin{equation}\label{eq:test}
\norme{e_r(p^n)}{\infty} \le \e \qquad |e_r(S_f^n)| \le \e \ .
\end{equation}
If the two inequalities (\ref{eq:test}) hold true, for $n = 1, 2, \dots, N$, then we can accept the numerical solution computed on the grid defined by $J_{g+1}$ and $N_{g+1}$, otherwise we have to increase these two integers and repeat the calculations.

\begin{figure}[!t]
\begin{picture}(300,200)(0,0)
\put(1,1){\resizebox{9cm}{!}{\includegraphics{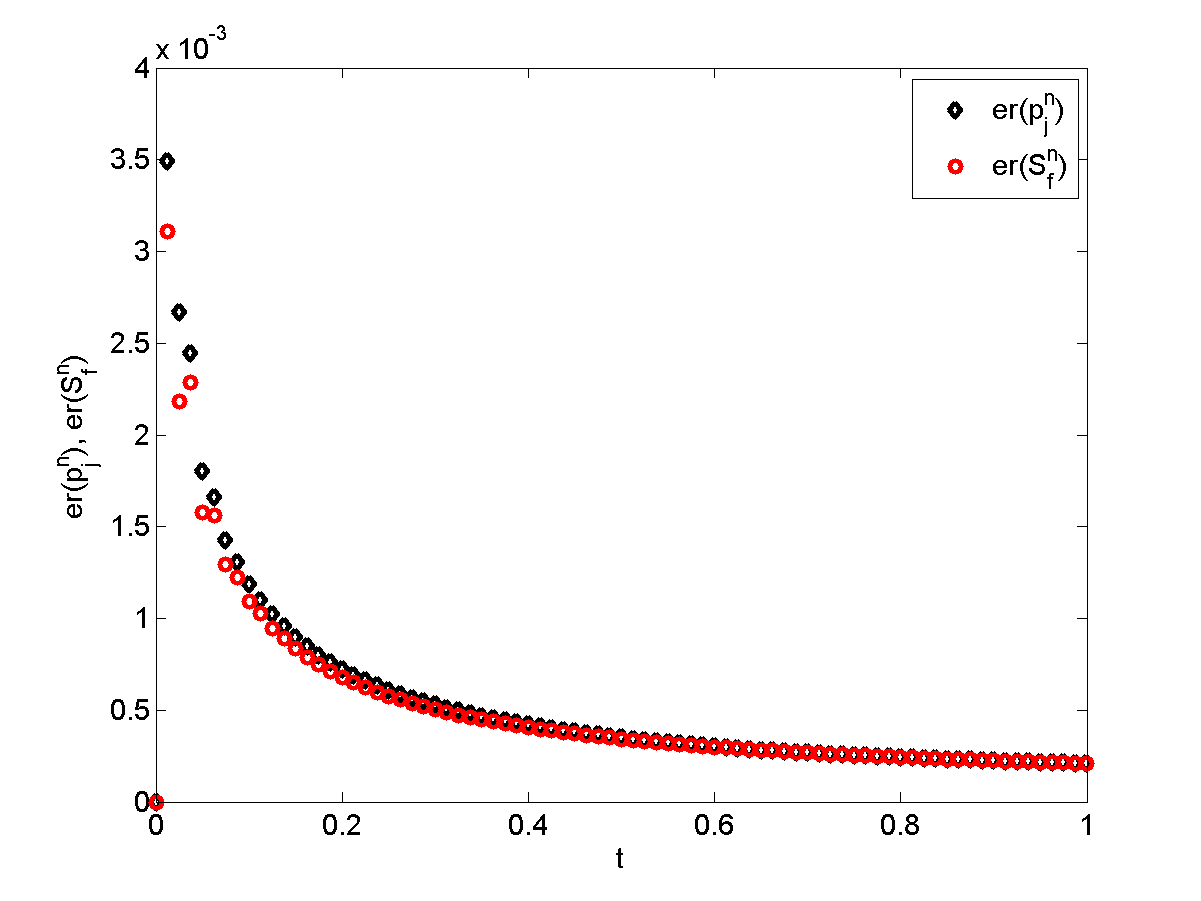}}}
\end{picture}
\vspace{-.3cm}
\caption{\it Numerical estimated errors $e_r(p^n)$ and $e_r(S_f^n)$ versus $t_n$.}
\label{fig:APOe}
\end{figure}

Fig.~\ref{fig:APOe} shows the error estimator results computed by setting $\e = 0.005$.
We fixed $\mu=20$ and started with $J_0=5$ and $J_1= 10$ repeating the computation by doubling the number of spatial grid-intervals if the required accuracy was not achieved. 
Our algorithm stopped when $J_4 = 80$ that for $\mu = 20$ means $N_4=320$.

\begin{figure}[!t]
\begin{picture}(300,360)(0,0)
\put(1,181){\resizebox{9cm}{!}{\includegraphics{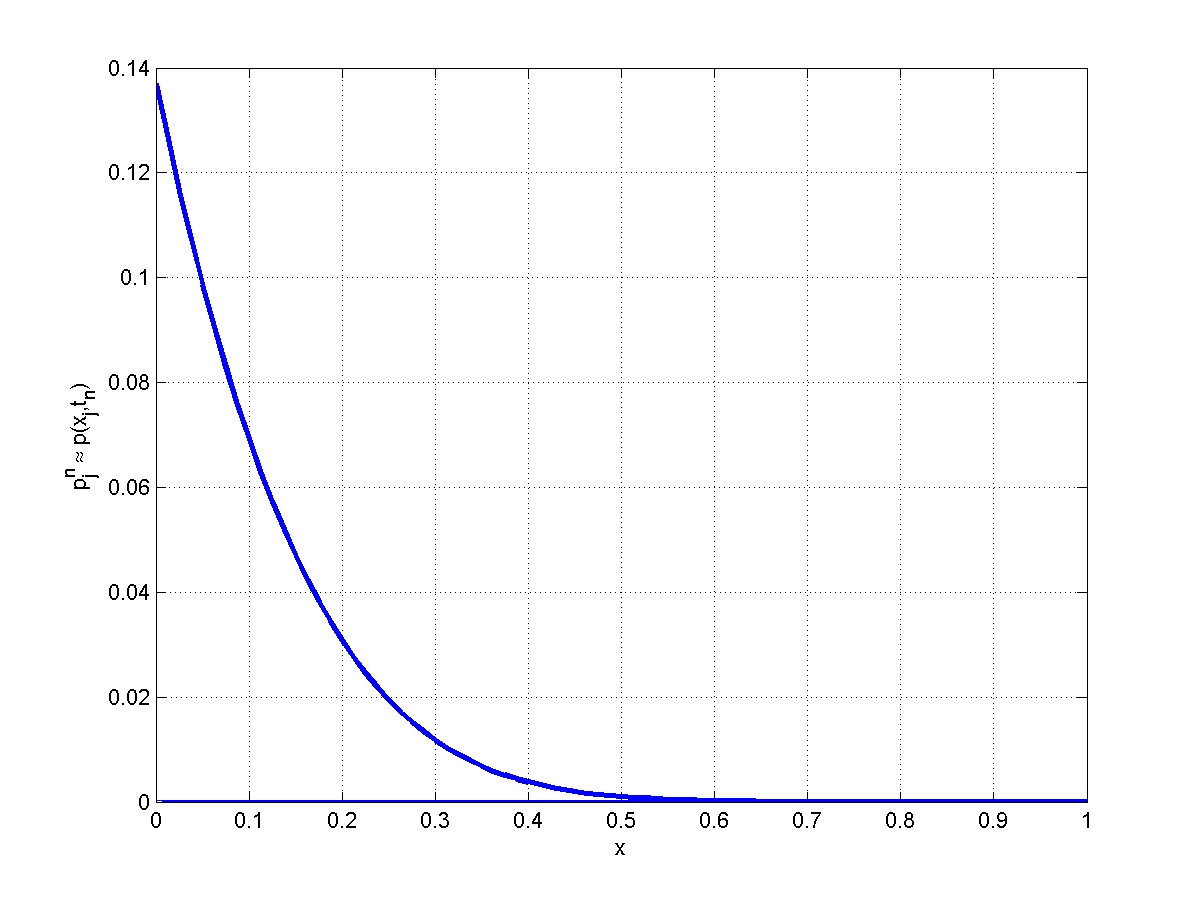}}}\\
\put(1,1){\resizebox{9cm}{!}{\includegraphics{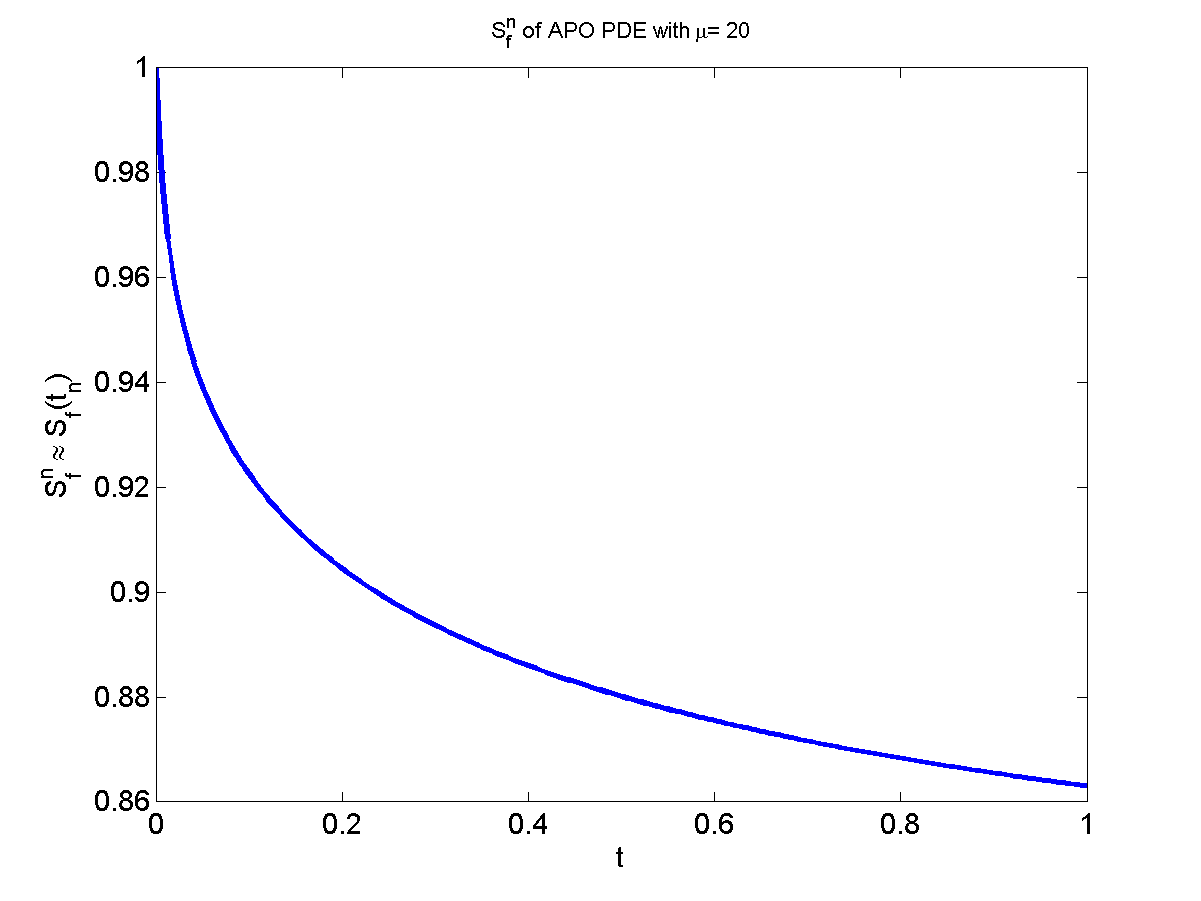}}}
\end{picture}
\vspace{-.3cm}
\caption{\it Numerical results: top frame $p_j^N$ versus $x_j$, bottom frame $S_f^n$ versus $t_n$.}
\label{fig:fdAPO}
\end{figure}

For the sake of completeness, in Fig.~\ref{fig:fdAPO} we plot $p_j^N$ versus $x_j$ and $S_f^n$ versus $t_n$, these results were obtained by the finite difference scheme with $N = 320$ and $\mu = 20$. 
From the numerical results shown in figure \ref{fig:APOe} we can easily realize that the greatest errors are found within a few time steps.
This suggests that a better accuracy can be achieved, without the use of Richardson's extrapolation, by developing an adaptive version of the explicit finite difference scheme.  

\appendix[The MATLAB script file]
Here we list the basic algorithm written in MATLAB. 
The script file is called \texttt{APOefds.m} and it can be easily modified to apply our a posteriori error estimator defined by equation (\ref{eq:est1}).
\begin{verbatim}
% APOefds.m
% APO equation on [0,xinf] by forward 
% in time and central in space
% J intervals in x, N intervals in t
% mi = k/h^2, k time step, h space step 
clear all; help APOefds
r = .1;     % parameters
T = 1;      % by Nielsen et al. 2002
sigma = .2;
E = 1;
xinf = 1;           
J = 80;
mi = 20
h = xinf/J
k = mi*h^2
N = ceil(T/k)
x = 0:h:xinf;
Stab1 =sigma^2/(abs(r-.5*sigma^2))-h
Stab2 =h^2/(sigma^2+r*h^2)- k
if (Stab1<0 | Stab2<0), break, end
w = zeros(J+1,1);          
s(1) = 1;
t(1)=0;
A = .5*mi*(sigma^2-(r-.5*sigma^2)*h);
B = 1-mi*sigma^2-r*k;
C = .5*mi*(sigma^2+(r-.5*sigma^2)*h);
A1 = 1+r*h^2/sigma^2;
B1 = 1+h+.5*h^2;
for n=1:N    %time loop
    dp1 = .5*(w(3,1)-w(1,1))/h;
    D = (A1-(A*w(1,1)+B*w(2,1)+ ...
		       C*w(3,1)-dp1))/(dp1+B1*s(n)); 
    s(n+1) = D*s(n);
    if (s(n+1)<0 | s(n+1)>1), break, end
    w(1,2) = 1-s(n+1);
    w(2,2) = A1-B1*s(n+1); 
    AM = A-.5*(s(n+1)-s(n))/(h*s(n));
    CM = C+.5*(s(n+1)-s(n))/(h*s(n));
    for j = 3:J    %space loop
         w(j,2) = AM*w(j-1,1)+ ...
				 B*w(j,1)+CM*w(j+1,1);
    end
    w(J+1,2) = 0;
    w(:,1) = w(:,2);
    t(n+1) = t(n)+k;
end
plot(t,s,'LineWidth',2.5)
axis([0 T .86 1]); grid
xlabel('t')
ylabel('S_f^n \approx S_f(t_n)')
\end{verbatim}



\ifCLASSOPTIONcaptionsoff
  \newpage
\fi



\end{document}